\pdfoutput=1
\documentclass[11pt, a4paper]{scrartcl}  
\usepackage[l2tabu, orthodox]{nag}  
\usepackage[utf8]{inputenc}
\usepackage[pdfusetitle]{hyperref}

\usepackage{amsmath, amsthm, amssymb}

\usepackage{mathptmx}
\usepackage[scaled=.90]{helvet}
\usepackage{courier}

\usepackage{graphicx}
\graphicspath{{figures/}} 
\usepackage{authblk}

\usepackage[semicolon, round]{natbib}
\usepackage{mathtools}  
\usepackage{gensymb}  
\usepackage[usenames]{color}  
\usepackage{cleveref}
\usepackage{listings}

\crefname{lstlisting}{listing}{listings}
\Crefname{lstlisting}{Listing}{Listings}

\numberwithin{equation}{section}
\numberwithin{table}{section}

\newcommand{\pp}[2]{\frac{\partial #1}{\partial #2}}

\begin{document}

\author[1,*]{Andrew~T.~T.~McRae}
\author[1]{Tim~N.~Palmer}
\affil[1]{Atmospheric, Oceanic and Planetary Physics, University of Oxford, Oxford, OX1 3PU}
\affil[*]{Correspondence to: \texttt{andrew.mcrae@physics.ox.ac.uk}}
\title{Using reduced-precision arithmetic in the adjoint model of MITgcm}
\date{}
\maketitle

\begin{abstract}
  In recent years, it has been convincingly shown that weather forecasting models can be run in single-precision arithmetic. Several models or components thereof have been tested with even lower precision than this. This previous work has largely focused on the main nonlinear `forward' model. A nonlinear model (in weather forecasting or otherwise) can have corresponding tangent linear and adjoint models, which are used in 4D variational data assimilation. The linearised models are plausibly far more sensitive to reductions in numerical precision since unbounded error growth can occur with no possibility of nonlinear saturation. In this paper, we present a geophysical experiment that makes use of an adjoint model to calculate sensitivities and perform optimisation. Using software emulation, we investigate the effect of degrading the numerical precision of the adjoint model. We find that reasonable results are obtained with as few as 10 significand bits, equal to the significand precision in the IEEE half-precision standard.
\end{abstract}

\section{Introduction}
\label{sec:intro}

Double-precision arithmetic, in which calculations are performed to some 16 decimal digits of accuracy, became standard in high-performance computing applications. In the past few years, there has been a surge of interest in running weather forecasting models at a more relaxed precision to make better use of computational resources. That this can be done without degrading forecast quality is unsurprising -- not only are initial condition errors present, but modern forecasting models explicitly include stochastic elements to better represent model uncertainty~\citep{palmer2009stochastic,leutbecher2017stochastic}. There is then little need for arithmetic calculations to be fully deterministic and exact to this level of accuracy~\citep{duben2014imprecise,palmer2014more}.

Some models have been tested entirely, or almost-entirely, at single precision: ECMWF's IFS~\citep{duben2014benchmark,vana2017single}, and MeteoSwiss's COSMO~\citep{rudisuhli2014cosmo}, which uses single precision operationally. The Met Office's UM uses single precision operationally in the main linear solve~\citep{maynard2019mixed}, and the dynamical core of the Japanese NICAM model has been tested at single precision~\citep{nakano2018single}. In an academic context, there have been several examples that use less-than-single-precision arithmetic with software emulation. This includes scale-selective precision experiments in a surface quasi-geostrophic model~\citep{thornes2018power} and in OpenIFS~\citep{chantry2018scale}, ensemble Kalman filter data assimilation experiments using SPEEDY~\citep{hatfield2018choosing}, and the NEMO and ROMS ocean models, the latter in 4D-Var mode~\citep{tinto2019mixed}.

With the exception of the recent work in ROMS, these examples have focused on the nonlinear `forward' model. Modern operational forecasting centres use variational data assimilation algorithms to obtain initial conditions for each forecast. These attempt to minimise a scalar cost-function which primarily represents the mismatch to observations. Gradient-based optimisation approaches are used, leading to the use of an adjoint model that computes the derivative of the cost function with respect to changes in the model state~\citep{ledimet1986variational,talagrand1987variational}. For efficiency reasons, an \emph{incremental} form of 4D-Var, as described at~\citet{courtier1994strategy}, has become standard in operational forecasting. This additionally requires a tangent-linear model.

It is plausible that these linearised models (the tangent-linear and adjoint models) react differently to a reduction in precision compared to the nonlinear model. For example, since they are linear, an instability can grow exponentially without any possibility of nonlinear saturation. Conversely, many operational forecasting models use simplified tangent-linear and adjoint models which are not exact linearisations of the nonlinear model. Since there is already an error in the gradient produced by the adjoint model, some mild reduction in numerical precision in the adjoint model can surely be tolerated.

In a related paper~\citep{hatfield2020single}, we consider a conjugate-gradient-based minimiser within an incremental 4D-Var setup, which more closely reflects modern data assimilation techniques. However, the underlying system used there is a simple quasi-geostrophic model. In this paper, we use a rather more complicated global ocean setup within MITgcm. However, our optimisation approach uses a more basic quasi-Newton minimiser and does not make use of the tangent-linear model, only the adjoint model. The remainder of the paper is laid out as follows: in \Cref{sec:prelim}, we give background information on MITgcm, adjoint models, and the software emulator of reduced precision that we use; in \Cref{sec:meth}, we discuss how we generate a reduced-precision adjoint model; in \Cref{sec:results} we present our numerical experiments, and in \Cref{sec:conc} we conclude and discuss future work.

\section{Preliminaries}
\label{sec:prelim}

\subsection{MITgcm}
MITgcm~\citep{marshall1997finite} is a software package that simulates atmospheric and oceanic processes using a finite-volume-based method. There are many options available for the discretisation of the underlying fluid equations, including a variety of timestepping schemes, advection schemes, different forms of the momentum equation, hydrostatic and non-hydrostatic modes, and so on. Several physics parameterisations are also available, including schemes for eddy parameterisation, vertical mixing, atmospheric physics, sea ice, and biogeochemistry. MITgcm is therefore considered an intermediate-complexity model.

A notable property of MITgcm, and the main reason we use it in this paper, is that corresponding adjoint models are also available. These have been used in sensitivity experiments~\citep{marotzke1999construction}, and particularly in the ECCO ocean state estimation project~\citep{stammer2002global,forget2015ecco}. The adjoint model is not stored as separate static code files (which would need to be updated whenever a change is made to the forward model, vastly complicating development). Rather, the adjoint model code is generated on-the-fly using automatic differentiation. The codebase is written to be compatible with the automatic differentiation tool TAMC, later TAF, as documented in~\citet{marotzke1999construction}. More recently, support was added for the free, open-source tool OpenAD~\citep{utke2008openadf}.

\subsection{Adjoints}
Formally speaking, a routine or program acts as a map from input variables $x_1, \ldots, x_n$ to output variables $y_1, \ldots, y_m$. For example, in the context of a GCM, we typically take the input variables $x_i$ to be components of the initial condition together with various parameters, and the $y_i$ to be the final state or statistics of interest. Occasionally, it is desirable to know not just the outputs, $y_i$, but also the derivatives of these with respect to (a subset of) the input variables, $\pp{y_i}{x_j}$. In a parameter tuning or data assimilation problem, it is common to construct a single output $J$, a \emph{cost function}, representing the mismatch between the evolution of the model and some observational data. The quantities $\pp{J}{x_i}$ are gradients of the cost function with respect to the initial condition and various parameters. This gradient can then be used to efficiently vary the parameters and/or initial conditions, depending on the application, in order to decrease the cost function $J$.

A natural way to compute the derivatives $\pp{y_i}{x_j}$ is to linearise each line of the model code around the (assumed nonlinear) model trajectory of the original program. This results in a tangent linear model, also known as forward mode differentiation. Given input variables $x_i$ and linear perturbations $\delta x_i$, this produces the linear output perturbation $\delta y_i$. To compute the derivatives $\pp{y_i}{x_j}$, this must be run once per input variable. While this is feasible for tuning a small number of parameters, it is impractical when the number of variables to be optimised over is large. Instead, a well-known technique is to use the adjoint method, also known as reverse mode differentiation. In the adjoint method, the derivatives are propagated backwards through the program from the final variables to the initial variables. The adjoint model must be run once per output variable. In the case of minimising a scalar cost function, the derivatives $\pp{J}{x_i}$ can therefore be found by running the adjoint model once, no matter how many input variables $x_i$ there are.

There are various technical difficulties with constructing an adjoint model. The key problem is that nonlinear variable quantities must be known in essentially the reverse order to which they are constructed by the original nonlinear program. For a toy program, it is possible to run the nonlinear program forward, saving all intermediate variables in memory, then running the adjoint model. However, for a GCM run, it is infeasible to store all intermediate variables across a large number of timesteps. The normal solution is \emph{checkpointing} -- regularly saving the model state to disk, so that only a small number of timesteps' worth of intermediate variables need to be kept in memory at once, and sophisticated algorithms exist that optimise the trade off between repeated execution and disk storage. The automatic differentiation software we use, OpenAD, includes the optimal binomial checkpointing algorithm introduced in Revolve~\citep{griewank2000revolve}.

Ultimately, adjoint models (and the far-simpler tangent linear models) can be built mechanically by processing code at an operation-by-operation level, and software frameworks exist that perform this in a highly-automated way. A thorough summary of the topic is given in~\citet{griewank2008evaluating}.

\subsection{Reduced-precision emulator}

The IEEE 754 double-precision floating-point format uses 64 bits to represent a number, with 1 sign bit, 11 exponent bits, and 52 significand bits. For typical bit patterns, the resulting number has the value
\begin{equation}
\pm ``1.b_{1} b_{2} \ldots b_\text{sbits}" \times 2^{e - \text{bias}},
\end{equation}
with a precision of around 16 decimal places. Here, sbits represents the number of significand bits, $b_{1} b_{2} \ldots b_\text{sbits}$ is the string of significand bits, and $e$ is the unsigned integer represented by the exponent bits (the total exponent is this non-negative integer minus a fixed bias of $2^{\text{ebits} - 1} - 1$, where ebits is the number of exponent bits). The single-precision format uses 32 bits -- 1 sign, 8 exponent, and 23 significand -- to represent numbers to a precision of around 7 decimal places. However, at present, typical hardware has no support for any precision between these or lower than single-precision, and nor does the Fortran programming language that many geophysical models are written in.

We therefore use a software emulator of reduced-precision arithmetic \citep{dawson2017rpe}. This provides reduced-precision Fortran variables embedded within double-precision numbers, with significand widths anywhere from 0 to 52 bits. The functionality is implemented as a Fortran data type \texttt{rpe\_var}, a derived type consisting of a \texttt{real} and an \texttt{integer}. Each \texttt{rpe\_var} maintains its floating-point value and a number of significand bits (which can therefore differ from variable to variable). Operator overloading is used, allowing normal \texttt{real} variables to be replaced by \texttt{rpe\_var} variables with minimal additional code changes required. Arithmetic operations are performed on the floating-point numbers natively, but the variables are truncated/rounded to the specified number of significand bits after every arithmetic operation. Unfortunately, this emulation of reduced-precision variables and arithmetic slows down code by one to two orders of magnitude. The experiments we perform are therefore only useful for analysing the numerical behaviour, rather than to realise improvements in computational performance.

\section{Methodology}
\label{sec:meth}

The main technical challenge is how to introduce reduced-precision variables into the automatically-generated adjoint code. The MITgcm adjoint build process both calls the automatic differentiation software and compiles the output code. The intermediate file is hundreds of thousands of lines long, and contains both nonlinear forward model code (with taping of nonlinear variables for use during the adjoint run) and the corresponding adjoint code. It is therefore undesirable to manually change this automatically-produced code in-place to introduce reduced-precision variables, as has been done for static codebases (e.g.\ \citet{chantry2018scale,hatfield2020single}), since it is hard to verify correctness. We instead use a simpler approach that lowers the precision of the adjoint model, but not in all variables.

Before explaining this in full, we first briefly explain how the automatic differentiation software OpenAD~\citep{utke2008openadf} interacts with a codebase, here MITgcm. Certain variables must be tagged by the user as \emph{independent} and \emph{dependent}; the tangent-linear or adjoint code then aims to compute the derivatives of the dependent variable(s) with respect to the independent variable(s). OpenAD parses the original code, and constructs a graph of the low-level arithmetic operations and the control flow. The tangent-linear and adjoint codes can be constructed by mechanically transforming this graph. OpenAD automatically identifies \emph{active} variables: these are the intermediate variables between the independent and dependent variables, plus the independent and dependent variables themselves.

For the active variables, both the original `nonlinear' values and the derivative values become variables in the program (the derivative part of an active variable has different meanings in tangent-linear and adjoint modes; typically, in tangent-linear (forward) mode, it is the derivative with respect to changes in some independent variable, while in adjoint (reverse) mode, it is the derivative of some dependent variable with respect to changes in that variable). Since there is a clear relationship between a nonlinear variable and its corresponding derivative variable (in both reverse and forward differentiation modes), it is common to relate these in code. The Tapenade AD tool~\citep{hascoet2013tapenade} uses a modified variable name, which is also common in manual codebases. OpenAD, however, changes the active variables to a Fortran derived type consisting of two reals. An active variable \texttt{foo} then has \texttt{foo\%v} representing the original value, and \texttt{foo\%d} representing the derivative component.

Our approach to introducing reduced precision to the adjoint is to modify the definition of the \texttt{active} derived type, changing just the derivative component from a double-precision \texttt{real} to an emulated \texttt{rpe\_var}. This means that the derivative variables are truncated to an appropriate precision whenever they are used, while the original nonlinear variables are unchanged. The nonlinear trajectory is therefore unaffected (since the aim of this study is to look at reduced precision in the adjoint). The drawback is that the nonlinear quantities enter the derivative calculations at full precision, which is not completely representative of a reduced-precision adjoint model. Since the nonlinear quantities are not just calculated once, but are periodically recomputed due to taping and checkpointing, changing the type of \texttt{foo\%v} would have other effects even if the emulator was inactive during the main forward run. As some extra precision leaks in, the results are plausibly slightly `optimistic' compared to if all variables in the adjoint model were at reduced precision. However, we argue that the effect is minimal: from basic theory, each primitive arithmetic operation involving active variables in the nonlinear code corresponds to at least one arithmetic operation involving derivative variables in the adjoint code. The number of reduced-precision calculations is therefore comparable to the total number of arithmetic operations. The code snippets \Cref{lst:pre} and \Cref{lst:post} show a representative calculation in the nonlinear forward model together with the corresponding adjoint code. The derivative variables are ubiquitous in the latter, and twice as many arithmetic calculations are performed.

\begin{lstlisting}[language=Fortran,basicstyle=\ttfamily\scriptsize,label={lst:pre},caption={Part of a nonlinear advection calculation in MITgcm.},captionpos=b]
 DO j=1-Oly+1,sNy+Oly
  DO i=1-Olx+1,sNx+Olx
   AdvectFluxVU(i,j) =
&  0.25*( vTrans(i,j) + vTrans(i-1,j) )
&      *( uFld(i,j) + uFld(i,j-1) )
  ENDDO
 ENDDO
\end{lstlisting}

\begin{lstlisting}[language=Fortran,basicstyle=\ttfamily\scriptsize,label={lst:post},caption={The automatically-generated adjoint code corresponding to \Cref{lst:pre}, showing that the \texttt{\%d} derivative variables are ubiquitous. In the innermost loop, eight arithmetic calculations are performed with derivative variables, compared to four arithmetic calculations in the nonlinear forward model.},captionpos=b]
J = 0+1*((22-0)/1)
do while (J.GE.0)
  I = 0+1*((47-0)/1)
  do while (I.GE.0)
    oad_dt_ptr = oad_dt_ptr-1
    OpenAD_Symbol_16129 = oad_dt(oad_dt_ptr)
    oad_dt_ptr = oad_dt_ptr-1
    OpenAD_Symbol_16130 = oad_dt(oad_dt_ptr)
    OpenAD_prp_328%d = OpenAD_prp_328%d+ADVECTFLUXVU(I,J)%d*(OpenAD_Symbol_16129)
    OpenAD_prp_329%d = OpenAD_prp_329%d+ADVECTFLUXVU(I,J)%d*(OpenAD_Symbol_16130)
    ADVECTFLUXVU(I,J)%d = 0.0d0
    UFLD(I,J+(-1))%d = UFLD(I,J+(-1))%d+OpenAD_prp_329%d
    UFLD(I,J)%d = UFLD(I,J)%d+OpenAD_prp_329%d
    OpenAD_prp_329%d = 0.0d0
    VTRANS(I+(-1),J)%d = VTRANS(I+(-1),J)%d+OpenAD_prp_328%d
    VTRANS(I,J)%d = VTRANS(I,J)%d+OpenAD_prp_328%d
    OpenAD_prp_328%d = 0.0d0
    I = I-1
  END DO
  J = J-1
END DO
\end{lstlisting}

\section{Experiment and Results}
\label{sec:results}

We use the \texttt{tutorial\_global\_oce\_optim} optimisation example supplied with MITgcm checkpoint 67m. A fuller description can be found in the online documentation~\citep{mitgcmdocs}, but we summarise the essentials here. This is an ocean-only experiment, with geography and bathymetry approximating that of the Earth. The grid is $4\degree \times 4\degree$, from 80$\degree$S to 80$\degree$N, with 15 vertical layers. The model is started from rest, and run for 1 year. Initial conditions for temperature and salinity are provided, as are wind stress and surface flux data.

The aim is to make the model temperature climatology $\bar{T}$ consistent with a provided dataset $\bar{T}^\text{obs}$ in the top two model layers. This is done by finding a (time-independent) surface heat flux adjustment $Q^\text{net,m}$, a 2D field, which is added to the provided heat-flux dataset. Specifically, we aim to minimise the cost function
\begin{equation}
  J = \lambda_1\cdot\underbrace{\frac{1}{N_1}\sum_{i=1}^{N_1}\left[\frac{\bar{T}_i - \bar{T}_i^\text{obs}}{\sigma_i^T}\right]^2}_\text{mismatch to observations} +\ \lambda_2\cdot\underbrace{\frac{1}{N_2}\sum_{i=1}^{N_2}\left[\frac{Q_i^\text{net,m}}{\sigma_i^Q}\right]^2}_\text{magnitude of adjustment},
\label{eq:costfn}
\end{equation}
where the $\bar{T}_i$ implicitly depend on the $Q_i^\text{net,m}$ according to the model evolution. The variables $\sigma_i^T$ and $\sigma_i^Q$ are provided uncertainty estimates for the temperature and surface heat flux variables. The number of $Q^\text{net,m}$ degrees of freedom, $N_2$, is 2315, while $N_1$ is approximately double this, and the parameters $\lambda_1$ and $\lambda_2$ are weights. This has the same form as a 4D-var cost function, having terms penalising both the mismatch to the observations and the adjustment to the control variables.

The adjoint model is used to compute the components of $\partial J / \partial Q^\text{net,m}$, the derivatives of the cost function with respect to the field $Q^\text{net,m}$. Starting with an initial guess $Q^\text{net,m} \equiv 0$ for the adjustment, a gradient-based optimisation scheme is used to produce an updated guess $Q^\text{net,m}$ that reduces $J$. This is repeated several times until the process is reasonably converged.

Compared to the default configuration, we reduced the various timestep parameters by a third, \emph{e.g.}, $\Delta T_{\text{mom}}$ = 1200s instead of 1800s. This is because the original configuration often shows an unstable checkerboard pattern in the South Pacific region when the gradient $\partial J / \partial Q^\text{net,m}$ is visualised at later iterations, even when double precision is used. This is then exacerbated by reducing numerical precision. We think it is reasonable to start from a stable configuration rather than one that is somewhat unstable, and so we reduce the timestep slightly.

In \Cref{ssec:sens}, we analyse the effect of a reduced-precision adjoint on the gradient $\partial J / \partial Q^\text{net,m}$ at the first iteration only. In \Cref{ssec:optim}, we then consider the full optimisation procedure.

\subsection{Effect of reduced precision on a single gradient}
\label{ssec:sens}

\begin{figure}[!ht]
\centering
\includegraphics[width=0.99\linewidth]{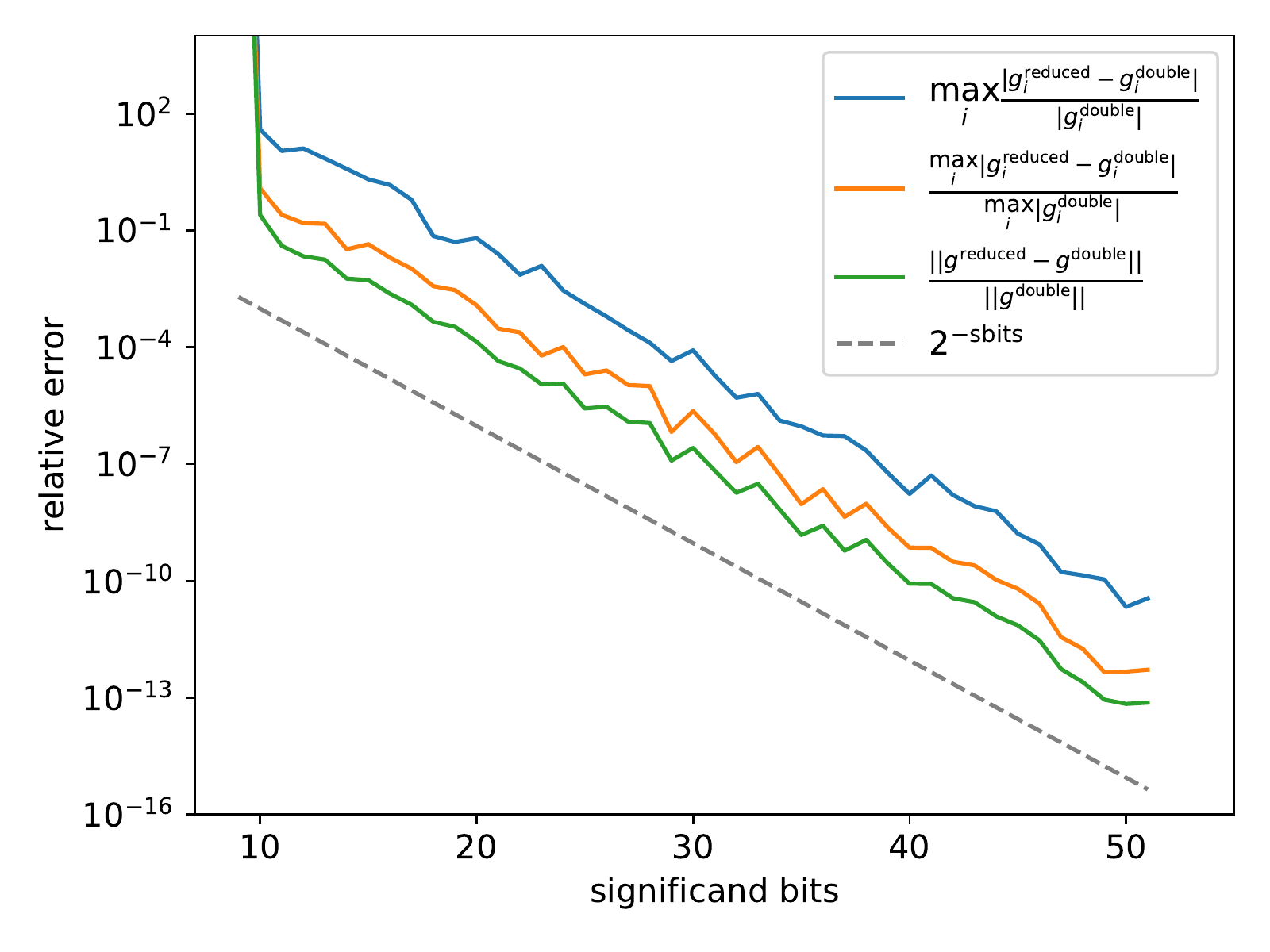}
\caption{This shows how $g \equiv \nabla J$, the gradient of $J$ with respect to the heat flux adjustment, changes as the numerical precision of the adjoint model is reduced. The orange and green lines represent the relative $l^\infty$ and $l^2$ errors of the entire gradient vector, respectively. The blue line represents the largest relative error in a single entry, although this is overly sensitive to very small entries in $(\nabla J)_i^\text{double}$. We see that these are all proportional to the dotted grey line, representing the machine epsilon at each precision level, until catastrophic blow-up occurs at 9 significand bits and below.}
\label{fig:gradgraph}
\end{figure}

The first run of the adjoint model produces $\nabla J$, the gradient of $J$ with respect to the $Q_i^\text{net,m}$, at the zero initial guess. Interpreted as a vector with 2315 components, \Cref{fig:gradgraph} shows how the computed gradient changes as the numerical precision of the adjoint model is reduced. As the number of significand bits is reduced from 52 (double precision), the gradient smoothly degrades down to 10 significand bits, the same as the IEEE half-precision standard. It is only at 9 significand bits and below that the gradient blows up drastically and becomes unusable.

\begin{figure}[!ht]
\centering
\includegraphics[width=\linewidth]{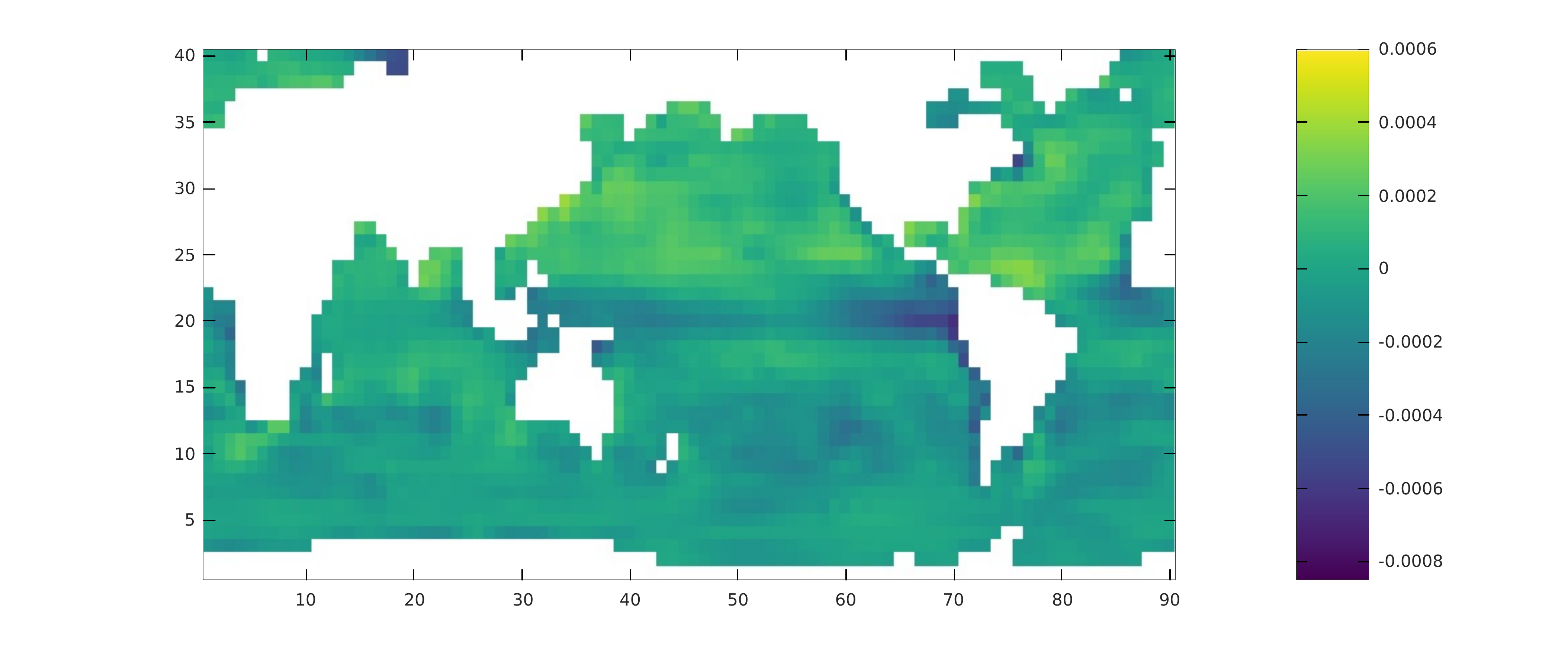}\\
\includegraphics[width=\linewidth]{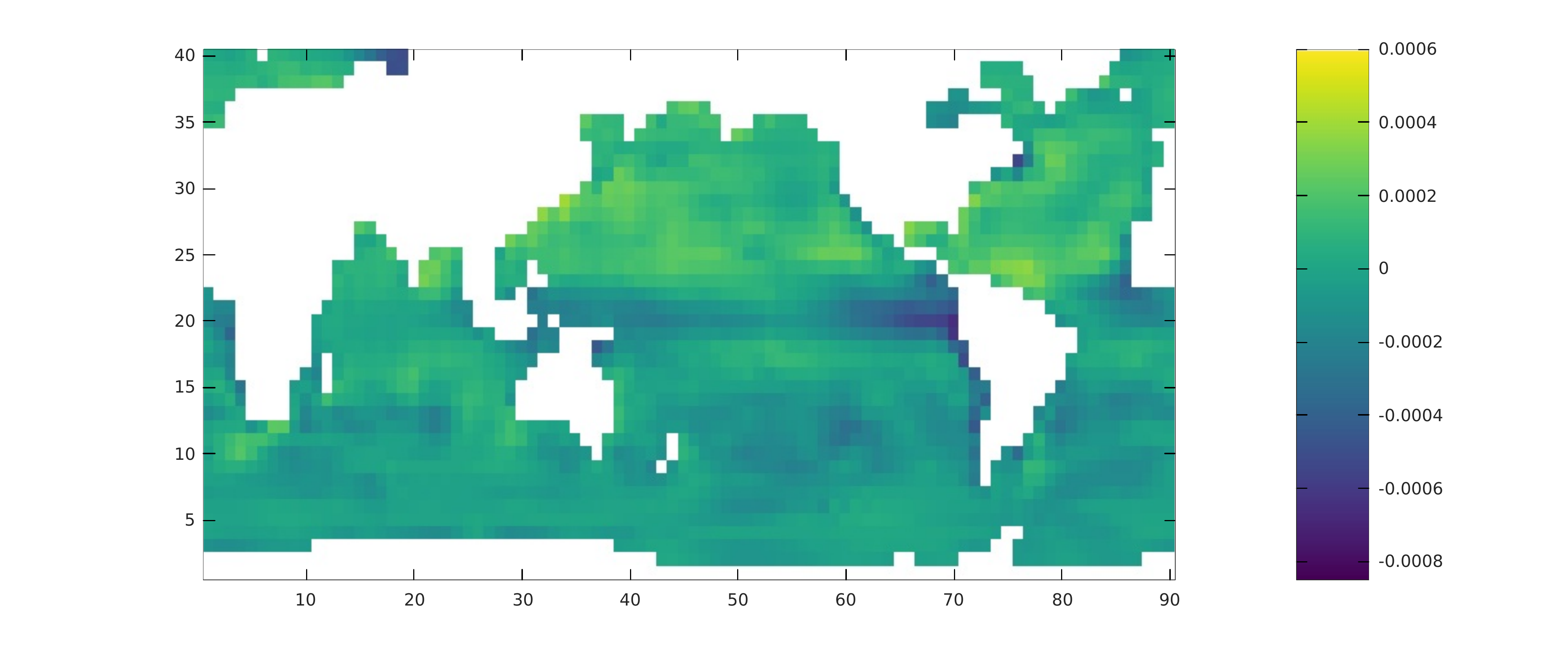}\\
\includegraphics[width=\linewidth]{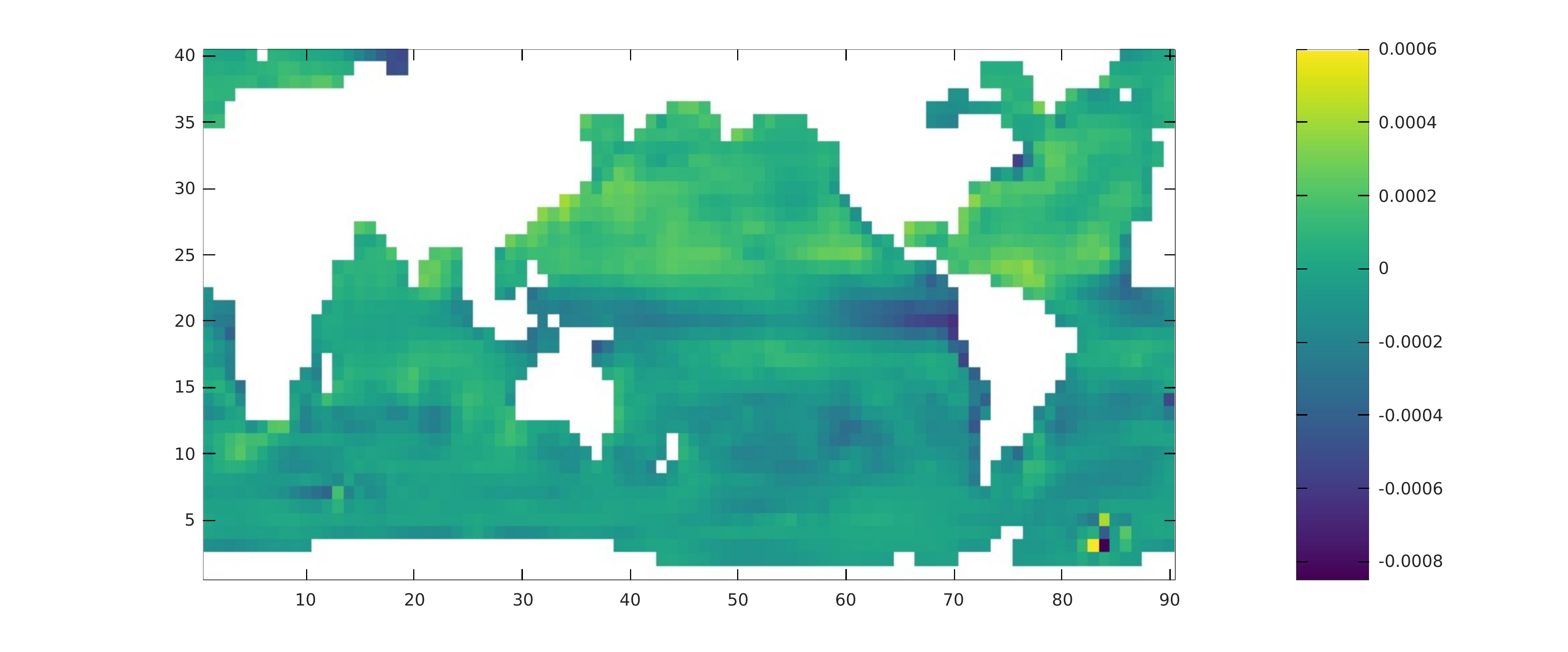}
\caption{A representation of $\nabla J$, the gradient of $J$ with respect to the heat flux adjustment at the first iteration, as a 2D field. Top: adjoint model at double-precision (52 significand bits). Middle: adjoint model using 11 significand bits. Bottom: adjoint model using 10 significand bits. The top two images are almost indistinguishable, implying that an adjoint model can produce acceptable gradients at heavily reduced numerical precision. At 10 bits, artifacts are visible in the south Atlantic and south of Madagascar, but the gradient is otherwise well-behaved.}
\label{fig:52vs11}
\end{figure}

In \Cref{fig:52vs11}, we show the gradient as a 2D field, as computed at double precision and at 11 and 10 significand bits. The first two pictures are nearly indistinguishable, and it is only on closer inspection that minor differences can be seen. At 10 bits, an instability is starting to develop in the south Atlantic region, but elsewhere the computed gradient is very similar.

\subsection{Effect of reduced precision on the full optimisation procedure}
\label{ssec:optim}

We now investigate to what extent a reduced-precision adjoint model affects the overall convergence of a gradient-based optimisation procedure. MITgcm includes an optimisation package \texttt{optim}/\texttt{lsopt}, but we instead use the `contributed' interface to M1QN3, a standard quasi-Newton package. Each iteration, M1QN3 is supplied with a new gradient $\nabla J$, from the adjoint model, and returns a new trial $Q^\text{net,m}$.

We modify some parameters slightly from their defaults. The main change is that, in the cost function \cref{eq:costfn}, we set $\lambda_1 = 1$ and $\lambda_2 = 0.4$. By default, these were set to 1 and 2. Our choice of parameters gives a \emph{harder} optimisation problem, since we are (relatively) penalising mismatches to the observations much more severely. The resulting problem is therefore less convex.

There are several parameters that we must set in the optimisation routine itself. We found that the original choice of parameters leads to some wasted iterations at double precision. Reduced-precision runs could then perform unnaturally well, or poorly. So that this does not lead to distracting results, we approximately optimised the initial step length for the double precision run by setting \texttt{dfminFrac} to 0.4. We then use this parameter for the reduced-precision runs.

\begin{figure}[!htb]
\centering
\includegraphics[width=0.75\linewidth]{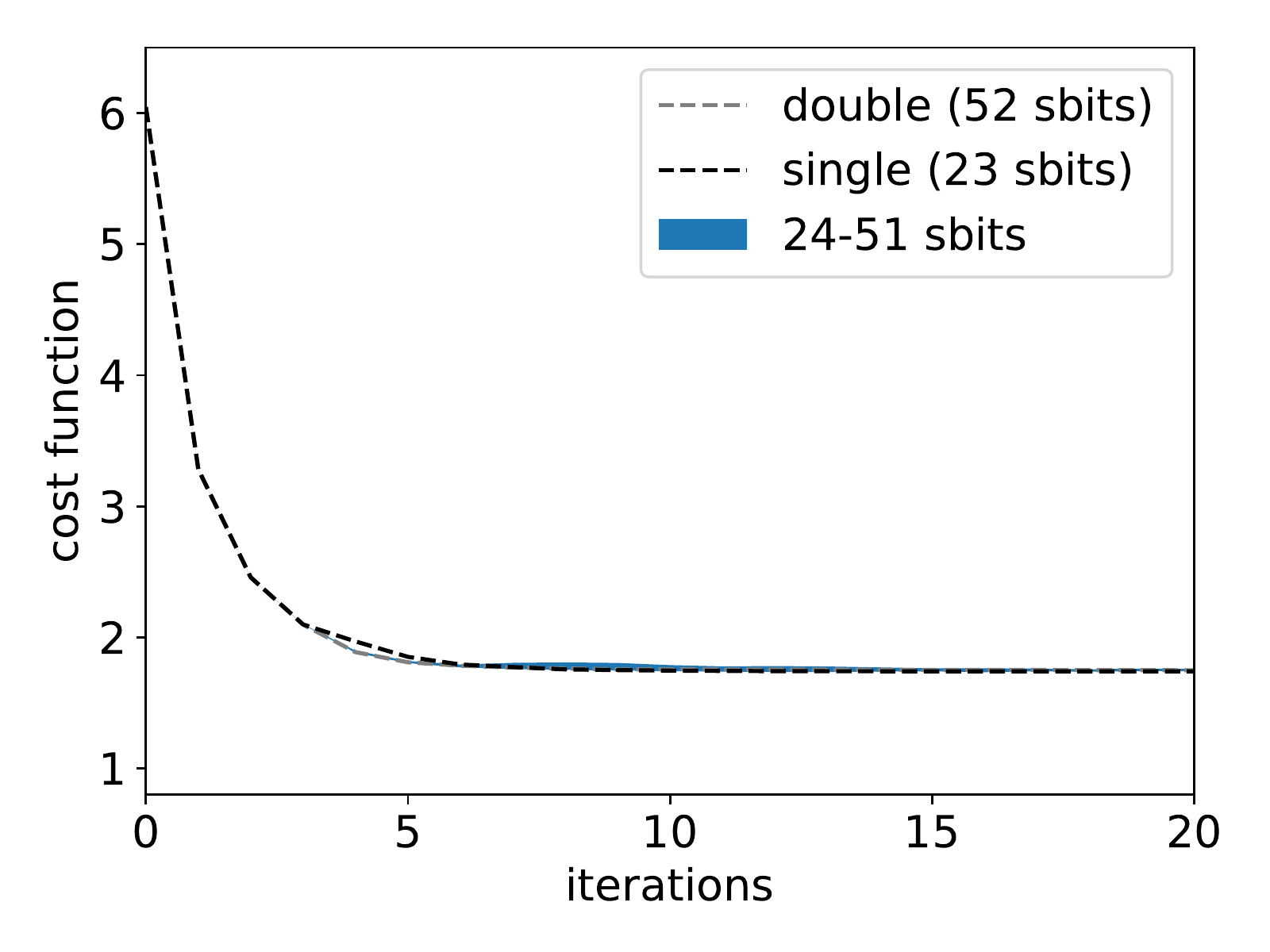}\\
\includegraphics[width=0.75\linewidth]{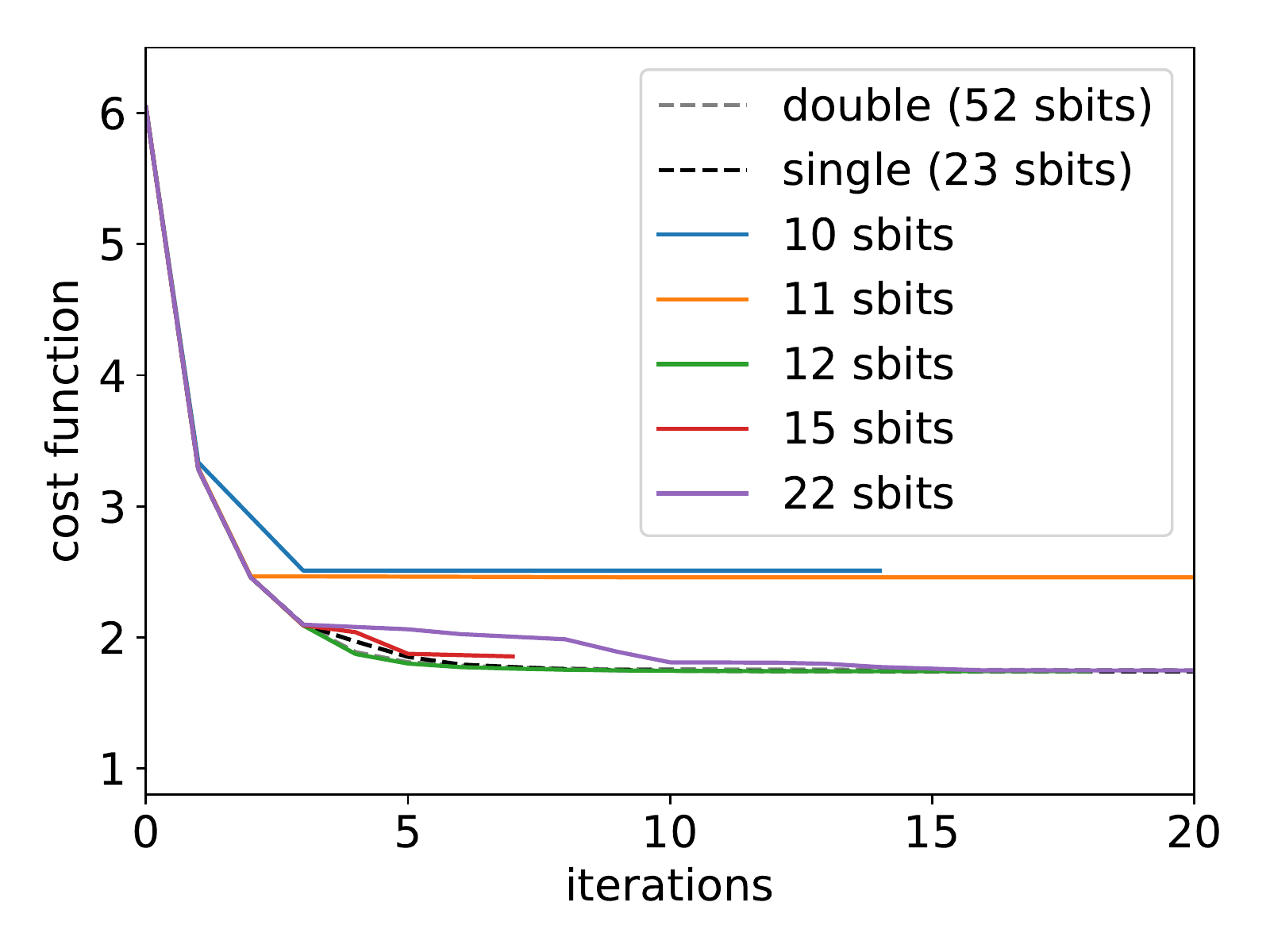}
\caption{Reduction in the cost function $J$ during the minimisation procedure, with the adjoint model run using varying numbers of significand bits from 52 (double-precision) down to 10 (half-precision). Top: optimisation using 23--52 significand bits. The (small!) shaded region shows the total variation for intermediate numbers of bits. Bottom: optimisation using less than 23 significand bits, selected values only. Until 16 significand bits, the optimisation procedure always works, although the convergence may be slowed (e.g., 22 sbits). Below this, the M1QN3 algorithm may by chance prematurely stall after some initial progress (e.g., 10, 11, and 15 sbits), or still succeed unhindered (e.g., 12 sbits). At 9 bits and below, the gradient is unusable and the optimisation procedure makes no progress; these are not shown.}
\label{fig:costfnconv}
\end{figure}

In \Cref{fig:costfnconv}, we show how the optimisation routine proceeds with various levels of reduced precision in the adjoint. At double precision, the cost function quickly decreases for 5--6 iterations, then makes little further progress. For various levels of reduced precision down to 16 significand bits, the convergence is similar. Below single precision -- 23 significand bits -- the optimisation procedure occasionally stagnates for a few iterations before making further progress. Note that the degradation is not strictly monotonic in the level of precision. For example, the 23 significand bit run converges slightly worse than the 12 significand bit run. At 15 significand bits and below, the optimisation algorithm often stalls early and makes no further progress, due to errors in the gradients and the previous steps that were taken. This could likely be repaired by restarting the M1QN3 algorithm with a basic gradient descent step, but we do not do that here.

\section{Conclusions and Outlook}
\label{sec:conc}

We have demonstrated that a heavily reduced-precision adjoint model is sufficient for convergence of a reasonably complicated geophysical optimisation problem. Of course, we are not claiming that adjoint models \emph{generally} produce better results at 12 significand bits than at 15, or anything so specific. Our results in the previous section should be interpreted as a single realisation of a potential ensemble of experiments, in which any perturbation to the forcing data or parameters would produce a different result. Averaged over a large number of such experiments, it is plausible that the behaviour would degrade smoothly as the number of significand bits is reduced, until a catastrophic blowup occurs. However, the exact threshold is sure to vary from model to model and likely also with resolution and timestep. A more detailed quantitative analysis of our specific scenario could be carried out, but its applicability to other models would be very limited. We note, however, that our results are reasonably consistent with~\citet{tinto2019mixed}, in which a 4D-Var setup in the ROMS ocean model is successfully performed with all variables at emulated single precision, and some 80\% of the variables at emulated half precision.

There are two natural directions in which this work could continue. One is to extend the work further towards state-of-the-art atmosphere and ocean models that are used for operational forecasting, attempting to run their tangent-linear and adjoint models in single precision or below. To become operationally useful, it would be necessary to transition away from software emulation towards native use of lower precision data types, which also have a reduced dynamic range. For single precision, this is unlikely to be a problem, but half precision only has a range of some $10^{-7}$ to $10^5$ and adhering to this would require much rewriting of code. Another direction is to identify components of the model which are more sensitive to reductions in precision, and possibly find workarounds. Such work has been performed in \citet{saffin2020inprep} in the context of forward-model parameterisation schemes, but it is unknown whether linearised and adjointed routines would behave the same way.

\section*{Acknowledgements}
The authors wish to thank Martin Losch for help with the M1QN3 optimisation package, and useful comments from Leo Saffin and Mat Chantry. This project has received funding from the European Research Council (ERC) under the European Union's Horizon 2020 research and innovation programme (ITHACA, an Information Theoretic Approach to Improving the Reliability of Weather and Climate Simulations, grant agreement no 741112).

\appendix

\section{Code availability}

All of the numerical experiments given in this paper were performed with the following versions of software, which we have archived on Zenodo: \citet{zenodo_mitgcm} and \citet{zenodo_m1qn3}.

\bibliography{mitgcmrpadj}

\begin{thebibliography}{30}
\providecommand{\natexlab}[1]{#1}
\providecommand{\url}[1]{\texttt{#1}}
\expandafter\ifx\csname urlstyle\endcsname\relax
  \providecommand{\doi}[1]{doi: #1}\else
  \providecommand{\doi}{doi: \begingroup \urlstyle{rm}\Url}\fi

\bibitem[Chantry et~al.(2018)Chantry, Thornes, Palmer, and
  D{\"{u}}ben]{chantry2018scale}
Matthew Chantry, Tobias Thornes, Tim Palmer, and Peter D{\"{u}}ben.
\newblock {Scale-Selective Precision for Weather and Climate Forecasting}.
\newblock \emph{Monthly Weather Review}, 147\penalty0 (2):\penalty0 645--655,
  2018.
\newblock \doi{10.1175/mwr-d-18-0308.1}.

\bibitem[Courtier et~al.(1994)Courtier, Th{\'{e}}paut, and
  Hollingsworth]{courtier1994strategy}
P.~Courtier, J.-N. Th{\'{e}}paut, and A.~Hollingsworth.
\newblock {A strategy for operational implementation of 4D-Var, using an
  incremental approach}.
\newblock \emph{Quarterly Journal of the Royal Meteorological Society},
  120\penalty0 (519):\penalty0 1367--1387, 1994.
\newblock \doi{10.1002/qj.49712051912}.

\bibitem[Dawson and D{\"{u}}ben(2017)]{dawson2017rpe}
Andrew Dawson and Peter~D. D{\"{u}}ben.
\newblock rpe v5: an emulator for reduced floating-point precision in large
  numerical simulations.
\newblock \emph{Geoscientific Model Development}, 10\penalty0 (6):\penalty0
  2221--2230, 2017.
\newblock \doi{10.5194/gmd-10-2221-2017}.

\bibitem[D{\"{u}}ben and Palmer(2014)]{duben2014benchmark}
Peter~D. D{\"{u}}ben and T.~N. Palmer.
\newblock {Benchmark Tests for Numerical Weather Forecasts on Inexact
  Hardware}.
\newblock \emph{Monthly Weather Review}, 142\penalty0 (10):\penalty0
  3809--3829, 2014.
\newblock \doi{10.1175/MWR-D-14-00110.1}.

\bibitem[D{\"{u}}ben et~al.(2014)D{\"{u}}ben, McNamara, and
  Palmer]{duben2014imprecise}
Peter~D. D{\"{u}}ben, Hugh McNamara, and T.N. Palmer.
\newblock {The use of imprecise processing to improve accuracy in weather {\&}
  climate prediction}.
\newblock \emph{Journal of Computational Physics}, 271:\penalty0 2--18, 2014.
\newblock \doi{10.1016/j.jcp.2013.10.042}.

\bibitem[Forget et~al.(2015)Forget, Campin, Heimbach, Hill, Ponte, and
  Wunsch]{forget2015ecco}
G.~Forget, J.-M. Campin, P.~Heimbach, C.~N. Hill, R.~M. Ponte, and C.~Wunsch.
\newblock {ECCO version 4: an integrated framework for non-linear inverse
  modeling and global ocean state estimation}.
\newblock \emph{Geoscientific Model Development}, 8\penalty0 (10):\penalty0
  3071--3104, 2015.
\newblock \doi{10.5194/gmd-8-3071-2015}.

\bibitem[Griewank and Walther(2000)]{griewank2000revolve}
Andreas Griewank and Andrea Walther.
\newblock {Algorithm 799: Revolve: An Implementation of Checkpointing for the
  Reverse or Adjoint Mode of Computational Differentiation}.
\newblock \emph{ACM Transactions on Mathematical Software}, 26\penalty0
  (1):\penalty0 19--45, 2000.
\newblock \doi{10.1145/347837.347846}.

\bibitem[Griewank and Walther(2008)]{griewank2008evaluating}
Andreas Griewank and Andrea Walther.
\newblock \emph{{Evaluating Derivatives: principles and techniques of
  algorithmic differentiation}}.
\newblock SIAM, Philadelphia, second edition, 2008.
\newblock ISBN 978-0-898716-59-7.

\bibitem[Hascoet and Pascual(2013)]{hascoet2013tapenade}
Laurent Hascoet and Val{\'{e}}rie Pascual.
\newblock {The Tapenade Automatic Differentiation Tool: Principles, Model, and
  Specification}.
\newblock \emph{ACM Transactions on Mathematical Software}, 39\penalty0
  (3):\penalty0 20:1--20:43, 2013.
\newblock \doi{10.1145/2450153.2450158}.

\bibitem[Hatfield et~al.(2018)Hatfield, D{\"{u}}ben, Chantry, Kondo, Miyoshi,
  and Palmer]{hatfield2018choosing}
Sam Hatfield, Peter D{\"{u}}ben, Matthew Chantry, Keiichi Kondo, Takemasa
  Miyoshi, and Tim Palmer.
\newblock {Choosing the Optimal Numerical Precision for Data Assimilation in
  the Presence of Model Error}.
\newblock \emph{Journal of Advances in Modeling Earth Systems}, 10\penalty0
  (9):\penalty0 2177--2191, 2018.
\newblock \doi{10.1029/2018MS001341}.

\bibitem[Hatfield et~al.(2020)Hatfield, McRae, Palmer, and
  D{\"{u}}ben]{hatfield2020single}
Sam Hatfield, Andrew McRae, Tim Palmer, and Peter D{\"{u}}ben.
\newblock {Single-precision in the tangent-linear and adjoint models of
  incremental 4D-Var}.
\newblock \emph{Monthly Weather Review, in press}, 2020.

\bibitem[{Le Dimet} and Talagrand(1986)]{ledimet1986variational}
Fran{\c{c}}ois-Xavier {Le Dimet} and Olivier Talagrand.
\newblock {Variational algorithms for analysis and assimilation of
  meteorological observations: theoretical aspects}.
\newblock \emph{Tellus A}, 38A\penalty0 (2):\penalty0 97--110, 1986.
\newblock \doi{10.3402/tellusa.v38i2.11706}.

\bibitem[Leutbecher et~al.(2017)Leutbecher, Lock, Ollinaho, Lang, Balsamo,
  Bechtold, Bonavita, Christensen, Diamantakis, Dutra, English, Fisher, Forbes,
  Goddard, Haiden, Hogan, Juricke, Lawrence, MacLeod, Magnusson, Malardel,
  Massart, Sandu, Smolarkiewicz, Subramanian, Vitart, Wedi, and
  Weisheimer]{leutbecher2017stochastic}
Martin Leutbecher, Sarah-Jane Lock, Pirkka Ollinaho, Simon T.~K. Lang,
  Gianpaolo Balsamo, Peter Bechtold, Massimo Bonavita, Hannah~M. Christensen,
  Michail Diamantakis, Emanuel Dutra, Stephen English, Michael Fisher,
  Richard~M. Forbes, Jacqueline Goddard, Thomas Haiden, Robin~J. Hogan, Stephan
  Juricke, Heather Lawrence, Dave MacLeod, Linus Magnusson, Sylvie Malardel,
  Sebastien Massart, Irina Sandu, Piotr~K. Smolarkiewicz, Aneesh Subramanian,
  Fr{\'{e}}d{\'{e}}ric Vitart, Nils Wedi, and Antje Weisheimer.
\newblock {Stochastic representations of model uncertainties at ECMWF: state of
  the art and future vision}.
\newblock \emph{Quarterly Journal of the Royal Meteorological Society},
  143\penalty0 (707):\penalty0 2315--2339, 2017.
\newblock \doi{10.1002/qj.3094}.

\bibitem[Marotzke et~al.(1999)Marotzke, Giering, Zhang, Stammer, Hill, and
  Lee]{marotzke1999construction}
Jochem Marotzke, Ralf Giering, Kate~Q. Zhang, Detlef Stammer, Chris Hill, and
  Tong Lee.
\newblock {Construction of the adjoint MIT ocean general circulation model and
  application to Atlantic heat transport sensitivity}.
\newblock \emph{Journal of Geophysical Research: Oceans}, 104\penalty0
  (C12):\penalty0 29529--29547, 1999.
\newblock \doi{10.1029/1999JC900236}.

\bibitem[Marshall et~al.(1997)Marshall, Adcroft, Hill, Perelman, and
  Heisey]{marshall1997finite}
John Marshall, Alistair Adcroft, Chris Hill, Lev Perelman, and Curt Heisey.
\newblock {A finite-volume, incompressible Navier Stokes model for studies of
  the ocean on parallel computers}.
\newblock \emph{Journal of Geophysical Research}, 102\penalty0 (C3):\penalty0
  5753--5766, 1997.
\newblock \doi{10.1029/96JC02775}.

\bibitem[Maynard and Walters(2019)]{maynard2019mixed}
C.M. Maynard and D.N. Walters.
\newblock {Mixed-precision arithmetic in the ENDGame dynamical core of the
  Unified Model, a numerical weather prediction and climate model code}.
\newblock \emph{Computer Physics Communications}, 2019.
\newblock \doi{10.1016/j.cpc.2019.07.002}.

\bibitem[{MITgcm}()]{zenodo_mitgcm}
{MITgcm}.
\newblock {Version of MITgcm used in 'Using reduced-precision arithmetic in the
  adjoint model of MITgcm'}, February 2020.
\newblock URL \url{https://doi.org/10.5281/zenodo.3636019}.

\bibitem[{MITgcm authors}()]{mitgcmdocs}
{MITgcm authors}.
\newblock {MITgcm Global Ocean State Estimation tutorial documentation}.
\newblock
  \url{https://mitgcm.readthedocs.io/en/latest/examples/global_oce_optim/global_oce_optim.html}.
\newblock Accessed: 2019-12-19.

\bibitem[Nakano et~al.(2018)Nakano, Yashiro, Kodama, and
  Tomita]{nakano2018single}
Masuo Nakano, Hisashi Yashiro, Chihiro Kodama, and Hirofumi Tomita.
\newblock {Single Precision in the Dynamical Core of a Nonhydrostatic Global
  Atmospheric Model: Evaluation Using a Baroclinic Wave Test Case}.
\newblock \emph{Monthly Weather Review}, 146\penalty0 (2):\penalty0 409--416,
  2018.
\newblock \doi{10.1175/MWR-D-17-0257.1}.

\bibitem[{optim\_m1qn3}()]{zenodo_m1qn3}
{optim\_m1qn3}.
\newblock {Version of optim\_m1qn3 used in 'Using reduced-precision arithmetic
  in the adjoint model of MITgcm'}, February 2020.
\newblock URL \url{https://doi.org/10.5281/zenodo.3635966}.

\bibitem[Palmer(2014)]{palmer2014more}
T.~N. Palmer.
\newblock {More reliable forecasts with less precise computations: a fast-track
  route to cloud-resolved weather and climate simulators?}
\newblock \emph{Philosophical Transactions of the Royal Society A:
  Mathematical, Physical and Engineering Sciences}, 372\penalty0
  (2018):\penalty0 1--14, 2014.
\newblock \doi{10.1098/rsta.2013.0391}.
\newblock URL
  \url{http://rsta.royalsocietypublishing.org/cgi/doi/10.1098/rsta.2013.0391}.

\bibitem[Palmer et~al.(2009)Palmer, Buizza, Doblas-Reyes, Jung, Leutbecher,
  Shutts, Steinheimer, and Weisheimer]{palmer2009stochastic}
T.N. Palmer, R.~Buizza, F.~Doblas-Reyes, T.~Jung, M.~Leutbecher, G.J. Shutts,
  M.~Steinheimer, and A.~Weisheimer.
\newblock {Stochastic Parametrization and Model Uncertainty}.
\newblock Technical report, 2009.
\newblock URL
  \url{https://www.ecmwf.int/en/elibrary/11577-stochastic-parametrization-and-model-uncertainty}.

\bibitem[R{\"{u}}dis{\"{u}}hli et~al.(2014)R{\"{u}}dis{\"{u}}hli, Walser, and
  Fuhrer]{rudisuhli2014cosmo}
Stefan R{\"{u}}dis{\"{u}}hli, Andr{\'{e}} Walser, and Oliver Fuhrer.
\newblock {COSMO in Single Precision}, 2014.
\newblock URL
  \url{http://www.cosmo-model.org/content/model/documentation/newsLetters/newsLetter14/cnl14{\_}09.pdf}.

\bibitem[Saffin et~al.(2020)Saffin, Hatfield, D{\"{u}}ben, and
  Palmer]{saffin2020inprep}
Leo Saffin, Sam Hatfield, Peter D{\"{u}}ben, and Tim Palmer.
\newblock {Reduced-precision parametrization: lessons from an
  intermediate-complexity atmospheric model}.
\newblock \emph{Submitted to Quarterly Journal of the Royal Meteorological
  Society}, 2020.

\bibitem[Stammer et~al.(2002)Stammer, Wunsch, Giering, Eckert, Heimbach,
  Marotzke, Adcroft, Hill, and Marshall]{stammer2002global}
D.~Stammer, C.~Wunsch, R.~Giering, C.~Eckert, P.~Heimbach, J.~Marotzke,
  A.~Adcroft, C.~N. Hill, and J.~Marshall.
\newblock {Global ocean circulation during 1992--1997, estimated from ocean
  observations and a general circulation model}.
\newblock \emph{Journal of Geophysical Research: Oceans}, 107\penalty0
  (C9):\penalty0 1:1--1:27, 2002.
\newblock \doi{10.1029/2001JC000888}.

\bibitem[Talagrand and Courtier(1987)]{talagrand1987variational}
Olivier Talagrand and Philippe Courtier.
\newblock {Variational assimilation of meteorological observations with the
  adjoint vorticity equation. I: Theory}.
\newblock \emph{Quarterly Journal of the Royal Meteorological Society},
  113\penalty0 (478):\penalty0 1311--1328, 1987.
\newblock \doi{10.1002/qj.49711347812}.

\bibitem[Thornes et~al.(2018)Thornes, D{\"{u}}ben, and
  Palmer]{thornes2018power}
Tobias Thornes, Peter D{\"{u}}ben, and Tim Palmer.
\newblock {A power law for reduced precision at small spatial scales:
  Experiments with an SQG model}.
\newblock \emph{Quarterly Journal of the Royal Meteorological Society},
  144\penalty0 (713):\penalty0 1179--1188, 2018.
\newblock \doi{10.1002/qj.3303}.

\bibitem[{Tint{\'{o}} Prims} et~al.(2019){Tint{\'{o}} Prims}, Acosta, Moore,
  Castrillo, Serradell, Cort{\'{e}}s, and Doblas-Reyes]{tinto2019mixed}
Oriol {Tint{\'{o}} Prims}, Mario~C. Acosta, Andrew~M. Moore, Miguel Castrillo,
  Kim Serradell, Ana Cort{\'{e}}s, and Francisco~J. Doblas-Reyes.
\newblock {How to use mixed precision in ocean models: exploring a potential
  reduction of numerical precision in NEMO 4.0 and ROMS 3.6}.
\newblock \emph{Geoscientific Model Development}, 12\penalty0 (7):\penalty0
  3135--3148, 2019.
\newblock \doi{10.5194/gmd-12-3135-2019}.

\bibitem[Utke et~al.(2008)Utke, Naumann, Fagan, Tallent, Strout, Heimbach,
  Hill, and Wunsch]{utke2008openadf}
Jean Utke, Uwe Naumann, Mike Fagan, Nathan Tallent, Michelle Strout, Patrick
  Heimbach, Chris Hill, and Carl Wunsch.
\newblock {OpenAD/F: A Modular, Open-Source Tool for Automatic Differentiation
  of Fortran Codes}.
\newblock \emph{ACM Transactions on Mathematical Software}, 34\penalty0
  (4):\penalty0 18:1--18:36, 2008.
\newblock \doi{10.1145/1377596.1377598}.

\bibitem[V{\'{a}}ňa et~al.(2017)V{\'{a}}ňa, D{\"{u}}ben, Lang, Palmer,
  Leutbecher, Salmond, and Carver]{vana2017single}
Filip V{\'{a}}ňa, Peter D{\"{u}}ben, Simon Lang, Tim Palmer, Martin
  Leutbecher, Deborah Salmond, and Glenn Carver.
\newblock {Single Precision in Weather Forecasting Models: An Evaluation with
  the IFS}.
\newblock \emph{Monthly Weather Review}, 145\penalty0 (2):\penalty0 495--502,
  2017.
\newblock \doi{10.1175/MWR-D-16-0228.1}.

\end{thebibliography}

\end{document}